\documentclass[prb,showpacs,twocolumn,amsmath,amssymb]{revtex4}
\usepackage{amssymb, amsmath}
\usepackage{graphicx}
\usepackage{float}
\usepackage{txfonts}
\usepackage{ulem}
\usepackage{bm}

\newcommand{\up}{\uparrow}
\newcommand{\dn}{\downarrow}

\usepackage{color}
\usepackage{ulem}
\begin{document}

\title{
%Two-dimensional Dirac-ring topological crystalline insulators\\
Topological insulating phases from two-dimensional nodal loop semimetals}

\author{
Linhu Li$^{1,2}$ and  Miguel A. N. Ara\'ujo$^{1,2,3}$
}

\affiliation{$^1$ Beijing Computational Science Research Center, Beijing 100089, China}

\affiliation{$^2$ CeFEMA, Instituto Superior
T\'ecnico, Universidade de Lisboa, Av. Rovisco Pais, 1049-001 Lisboa,  Portugal}

\affiliation{$^3$ Departamento de F\'{\i}sica,  Universidade de \'Evora, P-7000-671,
\'Evora, Portugal}

\begin{abstract}
Starting from a minimal model for a 2D nodal loop semimetal,
we study the effect of chiral mass gap terms.
%on the  corresponding "Dirac ring" Hamiltonian.
%we study the topological phases that arise when
%mass gap terms are introduced in the corresponding "Dirac ring" Hamiltonian.
The resulting
%We derive an expression for the Chern number
Dirac loop anomalous Hall insulator's Chern number
is the phase winding number of the mass gap terms on the loop.
We provide  simple lattice models,  analyze the topological
phases and generalize a previous index characterizing topological transitions.
The responses of the Dirac loop anomalous Hall
and quantum spin Hall insulators
to a magnetic field's vector potential
are also studied both in weak and strong field regimes, as well as the edge states
in a ribbon geometry.
\end{abstract}

\maketitle

\section{Introduction}

The nontrivial topological properties of fermions,
which have attracted great attention recently,
%Topological phases of fermions have attracted great attention recently.
%Their nontrivial properties
stem from their low energy Dirac-like band dispersion
and its associated chiralities.
Differently from conventional physical phases, topological phases
are classified by discrete topological invariants of occupied bands,
rather than continuous order parameters\cite{review1,review2}.
Depending on its time reversal, particle-hole and chiral symmetries,
a gapped system, insulator or  superconductor,
can be classified into ten topological classes, five of which can support
topologically nontrivial phases depending on the dimension of the system\cite{classes}.
In an insulating system, the bulk gap contains nontrivial boundary states
whose chirality or helicity is determined by the topological invariants.
In superconductors, the possibility of realizing Majorana fermions has
spurred intense research because of their potential application in
quantum computation\cite{alicea}.

In some three dimensional systems, there may also  be
linear band touching at discrete Dirac or Weyl
points, or ``nodes'', in the Brillouin Zone (BZ).
Dirac and  Weyl semimetals have been
the focus of intense research\cite{Wan,Delplace,Hou,Young,Ganeshan}, as
they are  gapless systems which can exhibit topological properties.
A Dirac semimetal enjoys both time-reversal and inversion symmetries.
When  one of these symmetries is broken, Weyl nodes
with opposite chiralities separated in momentum space may appear
and the semimetal exhibits surface  Fermi arcs and the chiral anomaly\cite{NN1}.
Examples of Dirac semimetals are
Na$_3$Bi, Cd$_3$As$_2$\cite{zawng,sxu,zklu,zwang2,zkliu,neupane,borisenko}.
The Weyl semimetal state has been experimentally confirmed in the
TaAs family\cite{hweng,shuang,sxu2,lv,lyang}.

More recently,
a new class of three-dimensional semimetal
with   nodal lines has attracted  growing interest\cite{Mullen,3D_loop,Yan},
following the suggestion for its realization in the hyperhoneycomb lattice\cite{Mullen}.
In this case,  the linear band touching occurs along a closed loop in the BZ.
The concept of nodal loop semimetal is relatively new and awaits further investigation.

In addition to the above types of three-dimensional topological semimetals
(Dirac, Weyl and nodal line),
a more recent proposal for
the concept of a {\it two-dimensional nodal line} semimetal has emerged and
a suggestion for its physical realization in a
new composite lattice composed of interpenetrating kagome and honeycomb lattices
has been presented\cite{Lu}.
Spin-orbit coupling can open a small gap at the node line,
resulting in a novel topological crystalline insulator.

Motivated by these recent developments,
we here study the  nodal loop (NL) semimetal in two dimensions,
for spinless fermions.
The introduction of mass gap terms may lead to topological insulating phases.
We derive an expression for the Chern number of the
resulting Dirac loop anomalous Hall insulator (DLAHI).
The Chern number is equivalent to the winding number of the mass terms' phase along the loop
and  can be regarded as the loop's chirality.
We examine the topological transitions that take place as model parameters
change and  generalize a previous index that characterizes such transitions.
The effect of a magnetic field on a DLAHI is also studied and compared
to the case of Dirac point systems.

In Section \ref{modelsec} we introduce the minimal model for a NL
semimetal, consider a mass gap,  and study the  topological properties of the DLAHI.
Section \ref{magsec} is devoted to the study of  magnetic field effects.
In Section \ref{conclsec} we summarize our results and
make some concluding remarks.

\section{Topological insulator in Generalized 2D nodal loop semimetal}\label{modelsec}
\subsection{Minimal Model and Topological Invariant}
\begin{figure}
\includegraphics[width=1\linewidth]{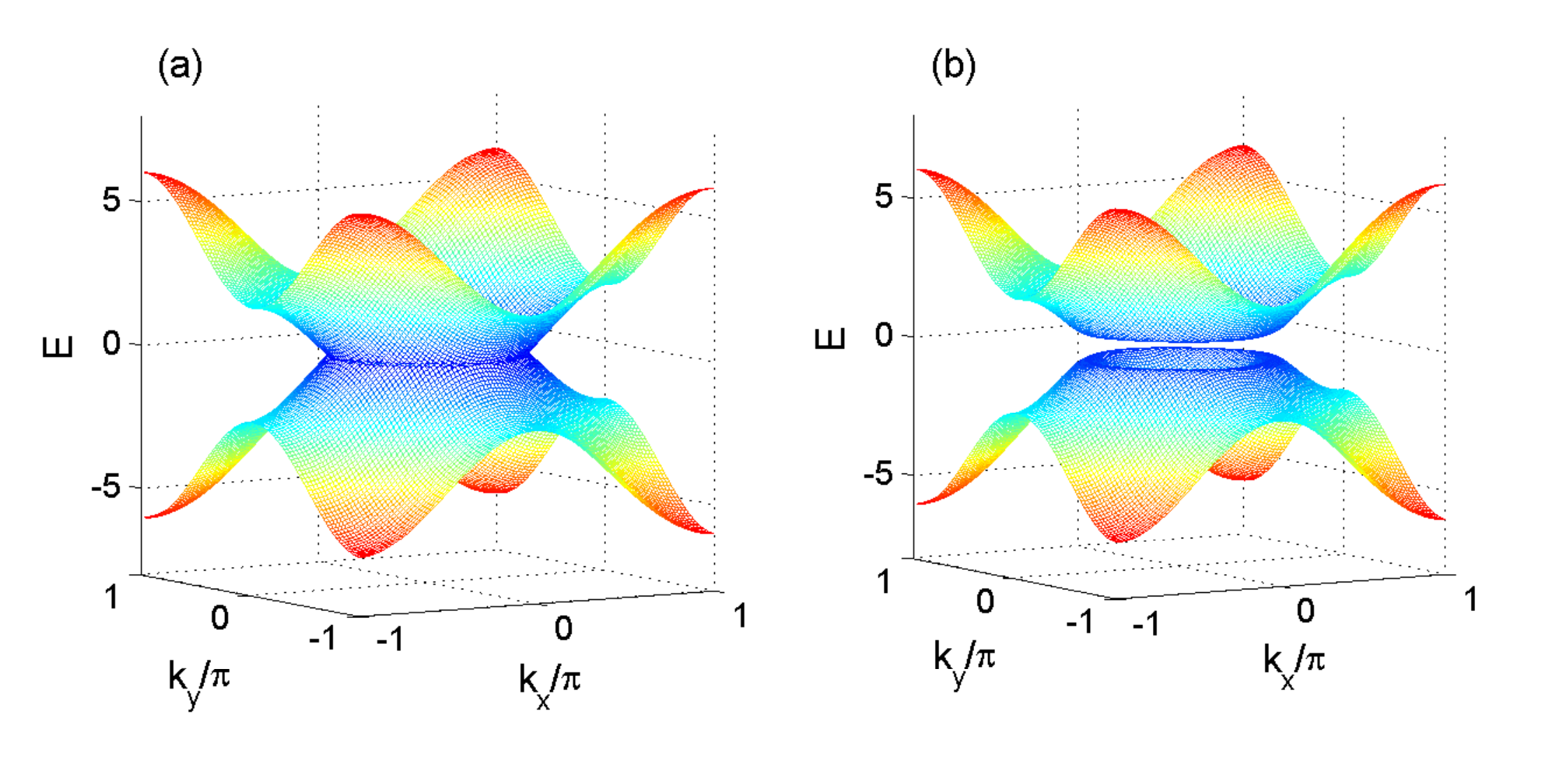}
\caption{(color online). The dispersion relation of a NL semimetal (a).
Panel (b) shows the gapped spectrum.
}\label{loopfig}
\end{figure}

To model a nodal loop semimetal in two dimensions,
it is necessary for the system to have at least two bands, and the Hamiltonian can be written as
\begin{eqnarray}
H=\bm{h}(\bm{k})\cdot\bm{\tau},\label{H}
\end{eqnarray}
where $\tau_{\alpha}$ ($\alpha=1,2,3$) are the Pauli matrices acting on sublattice
(``pseudo-spin'') space and
the Bloch wave vector $\bm{k}=(k_x,k_y)$ runs over the Brillouin zone (BZ).
We first consider a minimal Hamiltonian with a nodal circular loop\cite{3D_loop}:
\begin{eqnarray}
h_3=\hbar v_0 (k-k_0),\label{minimal}
\end{eqnarray}
Here $k=\sqrt{k_x^2+k_y^2}$ and $k_0$ is the loop radius.
Equation (\ref{minimal}) is supposed to be a valid
approximation in the region $k\approx k_0$ of the BZ.
This Hamiltonian gives a NL semimetal where the valence
and conduction bands cross at $k=k_0$
if $h_1=h_2=0$.
In a doped system, the Fermi surface would be a ring with radius $k_0$,
but we shall not consider doped systems below.
$h_1$ and $h_2$ can be viewed as two independent mass terms.
Nonzero $h_1$ and $h_2$ may open a gap
and turn the system into an insulator if
only the lower (valence) band is occupied.
We take the band gap $\Delta\equiv\sqrt{h_1^2+h_2^2}\ll \hbar v_0 k_0$.
It may take on a  constant value
on the loop, or have some $\bm k$ dependence.
A plot of the dispersion is shown in Fig. \ref{loopfig}.

%and holds two different P-type symmetries\cite{TopologicalClass1}: $\tau_1 h(k)\tau_1=-h(k)$ and $\tau_2 h(k)\tau_2=-h(k)$, each of them can be viewed as a chiral symmetry of this model. According to the tenfold way classification\cite{TopologicalClass1,TopologicalClass2}, this Hamiltonian belongs to either the AIII class (without time-reversal symmetry) or the BDI class (with time-reversal symmetry), which are both topologically trivial,

The topological properties of this model can be characterized
by the Chern number, $C$, of the occupied band, which is defined as
\begin{eqnarray}
C&=&\frac{1}{2\pi}\int V_{\bm{k}} dk_xdk_y\,,\label{Cformula}\\
V_{\bm{k}}&=&\partial_{k_x}A_{k_y}-\partial_{k_y}A_{k_x}\,,\label{Vk}
\end{eqnarray}
where $V_{\bm{k}}$ is the Berry curvature and
$A_{\alpha}= i\langle \varphi(\bm{k})|\partial_{\alpha}|\varphi(\bm{k})\rangle$
is the Berry connection\cite{Berry}, $\varphi(\bm{k})$ is a Bloch eigenstate of the
occupied  band, and the integral is over the two-dimensional Brillouin zone.
Equation (\ref{Vk}) yields a well defined result provided that  the loop is gapped.

We shall now show that a simple expression for $C$ can be obtained which involves
only the circulation of the phase of $h_1-ih_2$
along the loop.
For the two-band system described by Eq.(\ref{H}),
the Berry curvature of the lower band takes on the familiar form,
\begin{eqnarray}
V_{\bm{k}}&=&
\frac{1}{2|\bm h|^3}  \frac{\partial\boldsymbol h}{\partial k_x}
\times  \frac{\partial \boldsymbol h}{\partial k_y} \cdot \boldsymbol h\,,
%&=&\frac{1}{2|\bm h|^3}[\frac{\partial h_1}{\partial k_x}\frac{\partial h_2}{\partial k_y}h_3
%+\frac{\partial h_2}{\partial k_x}\frac{\partial h_3}{\partial k_y}h_1+
%\frac{\partial h_3}{\partial k_x}\frac{\partial h_1}{\partial k_y}h_2\nonumber\\
%&&-\frac{\partial h_2}{\partial k_x}\frac{\partial h_1}{\partial k_y}h_3-
%\frac{\partial h_1}{\partial k_x}\frac{\partial h_3}{\partial k_y}h_2-
%\frac{\partial h_3}{\partial k_x}\frac{\partial h_2}{\partial k_y}h_1]
\label{curvature}
\end{eqnarray}
where $|\bm h|=\sqrt{h_1^2+h_2^2+h_3^2}$.
 Using polar coordinates in momentum space, $(k,\theta)$, we write
$(k_x,k_y)=k(\cos{\theta}, \sin{\theta})$.
We rewrite the Berry curvature in polar coordinates and,
considering that the contribution to the integral (\ref{Cformula}) comes from the vicinity
of the loop where Eq.(\ref{minimal}) holds, we obtain
\begin{eqnarray}
C&=&\frac{\hbar v_0}{4\pi}\int \frac{dkd\theta}{|\bm h|^3}
\left(h_2\partial_{\theta} h_1-h_1\partial_{\theta} h_2\right)
\end{eqnarray}
The integration over $k$ can be performed under the assumption that
the gap $\Delta\ll \hbar v_0 k_0$. This allows us to extend the integration limits
of $k$ to the whole real axis and obtain:
\begin{eqnarray}
C &=& {\rm sgn}(v_0) \int_0^{2\pi} \frac{d\theta}{2\pi}
\frac{h_2\partial_{\theta} h_1-h_1\partial_{\theta} h_2}{h_1^2+h_2^2}\,,\label{winding}
\end{eqnarray}
where the integration is performed on the loop $h_3=0$.
The expression (\ref{winding}) is just the winding number for the phase of $h_1-ih_2$.
The above derivation may be regarded as an extension of the
contribution from a single Dirac point
to the Chern number\cite{eu}, which can be $\pm 1/2$.
By analogy, equation (\ref{winding}) assigns a chirality to the gapped loop.
While a fermionic system must have an even number of Dirac points\cite{NN}, there
 may be only one, or arbitrary number of NL's.

The above Chern number is ill defined when the gap closes, $h_1=h_2=0$,
as the integrand of Eq.(\ref{winding}) diverges.
At a topological phase transition the gap closes
and a definition of a $\mathbb{Z}$ index characterizing the transition
can be achieved in the extended three dimensional parameter space,  $(\bm{k},\eta)$,
where $\eta$ is a transition driving parameter\cite{Linhu}.
We assume the system to be an insulator for general $\eta$ and the gap closes at $\eta=\eta_0$.
If the gap closes at one or more discrete points in the BZ, a topological number
$C_p$ can be calculated
as the flux of the Berry curvature through a
sphere $\bm{S}$ enclosing each of these points in the
parameter space of $(\bm{k},\eta)$,
\begin{eqnarray}
C_p=\frac{1}{2\pi}\oiint \bm{V}\cdot d\bm{S}\,,\label{Cp}
\end{eqnarray}
where $\bm{V}= \nabla\times
i\langle \varphi(\bm{k},\eta)|\nabla|\varphi(\bm{k},\eta)\rangle$
is the Berry curvature in the extended parameter space.
The summation of $C_p$ over every gap closing point gives the change of
the Chern number $C$ across the transition\cite{Linhu}.

\begin{figure}
\includegraphics[width=1\linewidth]{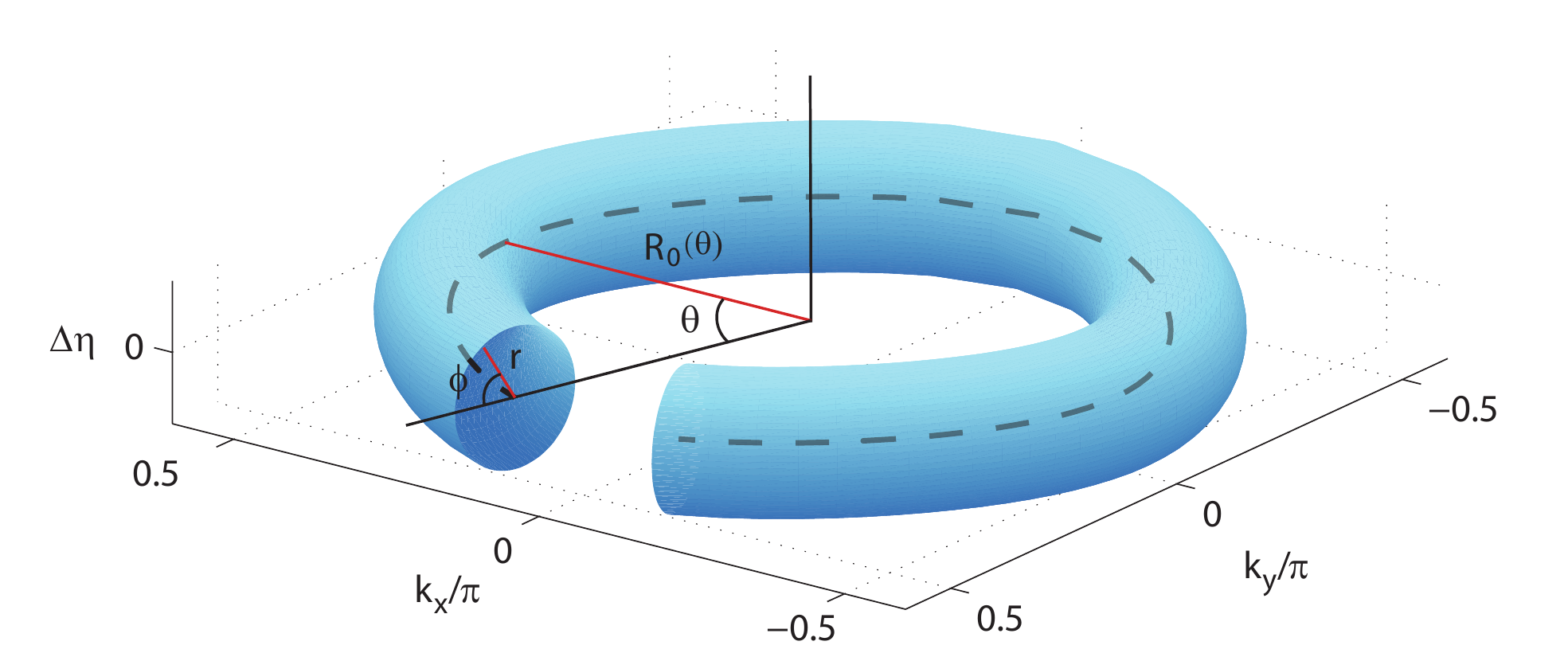}
\caption{(color online). The torus in Eq.(\ref{k_eta})
in the parameter space  $(\mathbf{k}, \Delta\eta\equiv\eta-\eta_0$.
The dashed line is the nodal loop when $\eta_0=0$.
The flux of $\bm{V}$ though the  torus surface gives the index $C_l$.}\label{Fig1}
\end{figure}

We can extend the above index to the case of a nodal loop semimetal, by defining
a similar index, $C_l$,
as the Berry curvature flux through a torus enclosing the loop.
In the parameter space of $\bm{k}$ and $\eta$,
this torus can be written as
\begin{eqnarray}
k_x&=&[k_0+r\cos{\phi}]\cos{\theta}\,,\nonumber\\
k_y&=&[k_0+r\cos{\phi}]\sin{\theta}\,,\nonumber\\
\eta&=&\eta_0+r\sin{\phi}\,,\label{k_eta}
\end{eqnarray}
where $\phi$ is a new  angular parameter on the torus and $r$ is the tube radius
as shown in Fig.\ref{Fig1}.
We may recast the surface integral in Eq.(\ref{Cp}) using equations (\ref{k_eta})
and $\theta$, $\phi$ as  integration variables, as
\begin{eqnarray}
C_l&=&\frac{1}{2\pi}\int\int V_{\phi,\theta} d\phi d\theta\,, \nonumber\\
V_{\phi,\theta}&=&\partial_{\phi}A_{\theta}-\partial_{\theta}A_{\phi}\,.\label{Cl}
\end{eqnarray}
The value of $C_l$ is independent of $r$ as long as no gap
closing point exists other than the nodal loop within the torus.
This index, $C_l$,  can serve as a topological invariant
which gives the change of Chern number, $C$, at a transition where  $\eta=\eta_0$.

\subsection{Specific models}

We now provide some specific lattice models to illustrate the topological phases of
a 2D nodal loop.
Note that in a lattice model, the nodal loop is not a perfect circle in the BZ,
so the loop radius $k_0$ is actually a function of the polar coordinate $\theta$,
$k_0=k_0(\theta)$.
Chern numbers for these lattice models are calculated numerically
with the method described in Ref.\cite{ChernNum}. Following this technique,
the parameter space $(k_x,k_y)$, or $(\theta,\phi)$ for the calculation of $C_l$,
is discretized and the circulations of the Berry connection on small plaquettes is performed.

We first consider the following  model
for the  vector $\bm{h}(\bm{k})$ in Eq.(\ref{H}):
\begin{eqnarray}
h_1(\bm{k})&=&\lambda\sin{k_y}+M\nonumber\\
h_2(\bm{k})&=&-\lambda\sin{k_x}\nonumber\\
h_3(\bm{k})&=&\mu-2(\cos{k_x}+\cos{k_y}).\label{model1}
\end{eqnarray}
The condition $h_3=0$ gives a nodal loop around the origin
for $0<\mu<4$, or around the point $(\pi, \pi)$ for $-4<\mu<0$.
We shall take  $\mu=2$ below.
When $M=0$, this model is also known as the Qi-Wu-Zhang model
for spin-1/2 systems\cite{QWZ}.
The term $M$ couples two pseudospin components at the same lattice site.

\begin{figure}
\includegraphics[width=1\linewidth]{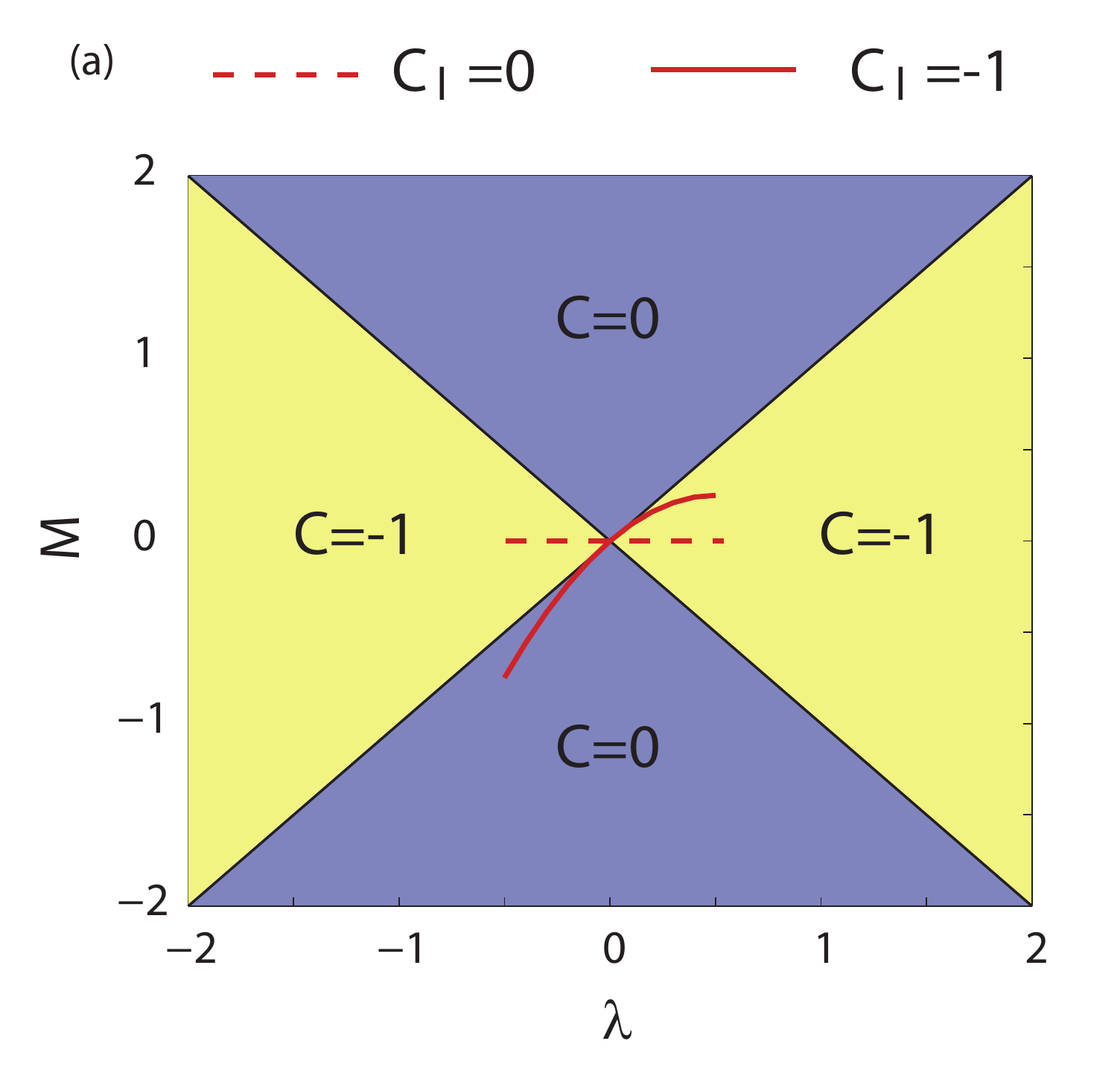}
\includegraphics[width=1\linewidth]{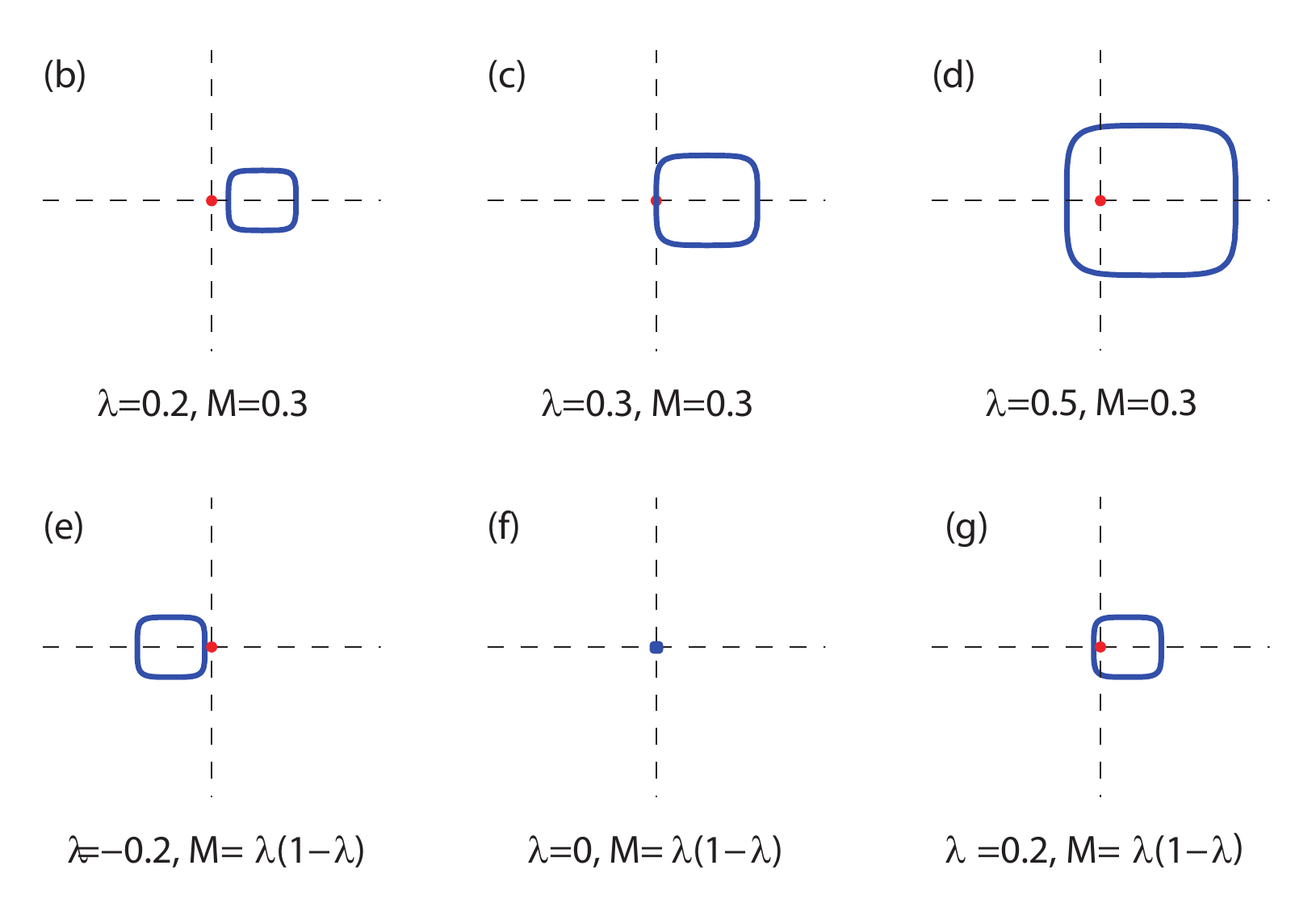}
\caption{(color online). Panel (a) shows the  phase diagram of model (\ref{model1}).
The red and dashed lines are trajectories along which $C_l$ is calculated,
as explained in the  text.
Panels (b)-(g) show  winding paths, $(h_1(\theta),-h_2(\theta))$, as $0\leq\theta<2\pi$
and $h_3=0$.
The direction of the winding path is clockwise for each case,
yielding $C=-1$ when it encloses the origin,
according to equation (\ref{winding}).}\label{Fig2}
\end{figure}

Model (\ref{model1}), for $4>\mu>0$, has two different topologically phases with $C=0$ or $C=-1$.
In Fig. \ref{Fig2}(a) we show the phase diagram with $\mu=2$. The topological phase boundary
is given by $|\lambda|=M$, where the gap closes at a single point
$\bm{k}=(0,-\mathrm{sgn}(\lambda\cdot M)\pi/2)$,
except in the case $\lambda=M=0$, where the system is a nodal loop semimetal.
In Figs. \ref{Fig2}(b)-(g) we plot  $(h_1(\theta),-h_2(\theta))$ for $h_3=0$, with
$\theta$ varying from $0$ to $2\pi$, for different parameter choices.
The Chern number $C=-1$ when $h_1-ih_2$ winds clockwise around the origin,
in accordance with equation (\ref{winding}).

We now study the topological transitions that take place
as the parameters $\lambda$ and $M$ change, either independently or
along a chosen curve in the $(\lambda,M)$ plane.
We start by examining two cases where
the spectral gap closes over the whole loop, at the transition.
A case where $M=0$ and $\lambda$
varies is plotted as a red dashed line in Fig. \ref{Fig2}(a). Going along such a
trajectory, no topological phase transition exists, as $C=-1$ always.
Now consider the case where $M=\lambda\left( 1-\lambda\right)$, which is plotted
as the red curve in Fig. \ref{Fig2}(a):
a transition between $C=0$ and $C=-1$ phases occurs at $M=\lambda=0$.
We use equation (\ref{Cl}) to calculate the index $C_l$, where we identify
the driving parameter $\eta$ with $\lambda$
and take the torus inner radius $r=10^{-4}$ in Eq.(\ref{k_eta}).
Equation  (\ref{Cl}) then  gives $C_l=0$ and $C_l=-1$
for these two cases, respectively,
which indicates the change of $C$ at the transition point.

In Figs. \ref{Fig2}(b)-(d) we show how the winding path
 $(h_1(\theta),-h_2(\theta))$ evolves for a topological phase
transition where the gap closes at only one point on the loop,
and the system evolves from $C=0$ to $C=-1$.
For this case, equation (\ref{Cp}) gives $C_p=-\mathrm{sgn}(\lambda\cdot M)$, as expected.
Similarly,  Figs. \ref{Fig2}(e)-(g) show
the winding paths of  $h_1-ih_2$
as the model evolves with $-0.2\leq\lambda\leq 0.2$ and $M=\lambda\left( 1-\lambda\right)$.
For this case we obtain $C_l=-1$.
Note that Figs. \ref{Fig2}(c) and (f) show  winding paths for
two gapless spectra: panel (c) shows the situation where the gap closes at only one point of the loop;
panel (f) depicts the case where the gap vanishes over the whole loop.

\begin{figure}
\includegraphics[width=1\linewidth]{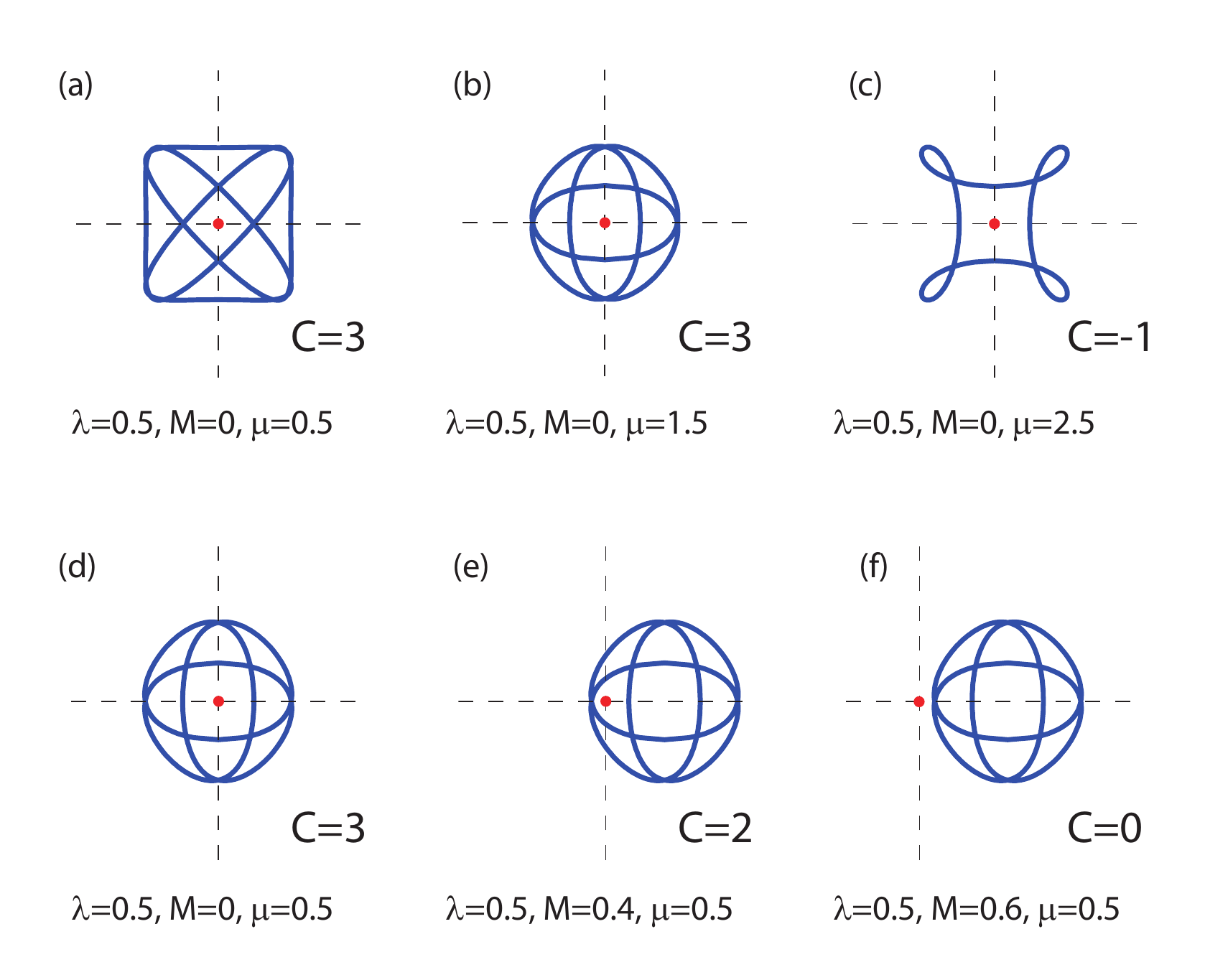}
\caption{(color online). Winding paths  $(h_1(\theta),-h_2(\theta))$
for model (\ref{model2}) for different parameter choices.
The direction of the winding path is anti-clockwise
except in (c), where  it is clockwise.}\label{Fig3}
\end{figure}

In the above lattice  model,
the Chern number may only be $\pm1$ or $0$,
as the winding path may go around the origin no more than once.
However, if the mass terms $h_1$ and $h_2$ contain higher harmonics,
the winding path will be more complicated
and the system may have higher Chern number phases.
As an example, we can choose the vector $\bm{h}(\bm{k})$ as
\begin{eqnarray}
h_1(\bm{k})&=&\lambda\sin(2k_y)+M\,,\nonumber\\
h_2(\bm{k})&=&-\lambda\sin(2k_x)\,,\nonumber\\
h_3(\bm{k})&=&\mu-2(\cos k_x +\cos k_y)\,.\label{model2}
\end{eqnarray}
In Fig.\ref{Fig3} we plot  $(h_1(\theta),-h_2(\theta))$
for different parameter choices together with  the corresponding Chern numbers.
Panels (a)-(c) illustrate  how  $\mu$ changes the shape of the winding paths.
and panels (d)-(f) show how
 $M$ changes the position of the winding path relative to the origin.
$\lambda$ will change the size of the path (not shown in the figures).

In the recent proposal\cite{Lu} for the realization of the 2D nodal loop system
in a kagome-honeycomb mixed lattice,
spin-orbit terms were shown to introduce mass gap terms with
$d_{xy}$ and $d_{x^2-y^2}$ symmetry.
The system considered is time-reversal invariant and such terms yield $C=\pm2$
in two effectively decoupled subspaces.

\section{Magnetic field}
\label{magsec}

We now take the nodal loop Hamiltonian (\ref{H})-(\ref{quadratic})
as a model for spinless fermions on
a lattice and analyze the effect of a magnetic field's vector potential.

In the case of
the anomalous Hall insulator with Dirac points, it has been shown that a weak field
turns the system into a metal with
half-filled Landau levels (LLs)\cite{haldane1988,Morais,canadianos,eu},
while a topologically trivial insulator remains insulating under the field.

In order to understand the effect of a weak magnetic field on a NL,
we follow the same procedure as in Ref\cite{eu} and introduce
a vector potential, $\boldsymbol A$,  which  minimally couples
to the orbital degrees of freedom,
$-i\hbar\nabla \rightarrow -i\hbar\nabla - e\boldsymbol A$,
where $e<0$ denotes the electron charge.
It is convenient to replace Eq.(\ref{minimal}) with
\begin{eqnarray}
h_3=\hbar^2\frac{k^2-k_0^2}{2m}\,, \label{quadratic}
\end{eqnarray}
which, when linearized yields Eq.(\ref{minimal}).
We rewrite (\ref{quadratic}) in real space as:
\begin{eqnarray}
h_3=-\hbar^2\frac{\partial_x^2+\partial_y^2+k_0^2}{2m}
\label{h3analytic}
\end{eqnarray}
If we further choose constant $h_{1(2)}$ with
$\sqrt{h_1^2 + h_2^2}\equiv \Delta \ll \frac{\hbar^2k_0^2}{2m}$, then the NL
becomes gapped is a topologically  trivial insulator at half-filling.
We write $\boldsymbol A=B(0,x,0)$ and the wave functions as:
\begin{eqnarray}
\psi_{\kappa,n}&=&e^{i\kappa y}\phi_n\left( x-\frac{\hbar \kappa}{eB} \right)
\cdot\left(\begin{array}{c} \alpha \\ \beta
\end{array}\right)\,.\label{LLsexpress}
\end{eqnarray}
Here,  $n=0,1,...$ denotes the LL index, $\omega_c=|eB|/m$ denotes the cyclotron frequency,
and $\phi_n(x)$ is a harmonic oscillator wave function.
The column vector $(\alpha ,\beta)^T$ solves the eigenproblem:
\begin{eqnarray}
\left\{
\left[\left(n+\frac 1 2 \right)\hbar\omega_c- \frac{(\hbar k_0)^2}{2m}\right] \tau_3
+ h_1\tau_1 + h_2\tau_2\right\} \left(\begin{array}{c} \alpha \\ \beta
\end{array}\right)=E_n\left(\begin{array}{c} \alpha \\ \beta
\end{array}\right) \,, \label{trivial matrix}
\end{eqnarray}
and the energy levels are given by:
\begin{eqnarray}
E_n&=&\pm \sqrt{ \left[\left(n+\frac 1 2 \right)\hbar\omega_c - \frac{(\hbar k_0)^2}{2m}
\right]^2 + \Delta^2
}\,.\label{LLstrivial}
\end{eqnarray}
The lowest LL ($n=0$) is far from the Fermi level if $\hbar\omega_c\ll (\hbar k_0)^2/(2m)$.
At half-filling the lower "-" branch LL's are fully occupied and the ones close to the Fermi level
have high index $n=N$ such that
$\left(N+1/2 \right)\hbar\omega_c \approx  (\hbar k_0)^2/(2m)$.

The charge Hall conductance, $\sigma_{yx}$, takes on quantized values between the LLs.
Each filled LL contributes $e^2/h$ to  $\sigma_{yx}$.
In the loop gap,  $\sigma_{yx}=0$.

Because high order LLs are close to Fermi level,
one may consider the
semiclassical dynamics in the magnetic field.
The semiclassical motion is determined by the equations:
\begin{eqnarray}
\hbar \dot{\bm k} = e v_{\bm k} \times \bm B\,,
\qquad
\dot{\bm r} =  v_{\bm k}\label{semicl}
\end{eqnarray}
where the group
velocity, $v_{\bm k}$,  in the upper/lower band obeys
$\hbar v_{\bm k}=\pm\partial |\bm h|/\partial \boldsymbol k$.
Within this approach the electrons follow orbits in the $(k_x,k_y)$ plane
determined by the Bohr-Sommerfeld quantization rule.
The equations (\ref{semicl})  cannot be directly applied to the gapped loop Hamiltonian,
however.
Instead, they may be applied to either branch of the massless loop in equation
(\ref{h3analytic}) with $h_1=h_2=0$. If we take $B>0$,  for instance, then
the electron orbits in $\bm k$-space  go counterclockwise
for the lower  branch, while they go clockwise in the upper 
branch. The mass terms $h_{1(2)}$ cause quantum mixing of the orbits
of both branches, as equation (\ref{trivial matrix}) explicitly shows.
As we go up in energy, we loose clockwise and gain counterclockwise orbits, hence
the Hall conductance increases.

Considering now the case of a topological system,  we take
$h_1=\hbar v k_y$,  $h_2=-\hbar v k_x$ and $h_3$ from equation (\ref{h3analytic}).
 Then the gap on the loop is
$\Delta= \hbar v k_0 \ll (\hbar k_0)^2/(2m)$. The Chern number of the lower band is $C=-1$.
It is convenient to define the operator
$\hat{\cal O}=ip_x+\hbar \kappa-eBx$ which obeys the commutation relation
$\left[ \hat{\cal O},\hat{\cal O}^\dagger  \right]=-2\hbar eB$.
The Hamiltonian now takes the form
\begin{eqnarray}
\hat H=
\left( \begin{array}{cc}
\frac{\hat{\cal O}^\dagger\hat{\cal O}-\hbar e B - (\hbar k_0)^2}{2m} & v \hat{\cal O}
\\
v \hat{\cal O}^\dagger & -\frac{\hat{\cal O}^\dagger\hat{\cal O}-\hbar e B - (\hbar k_0)^2}{2m}
\end{array}\right)
\end{eqnarray}
The eigenstates for $B>0$ involve the same set of harmonic oscillator functions $\phi_n$
above:
\begin{eqnarray}
\psi_{\kappa,n}=e^{i\kappa y}\left( \begin{array}{c}
\alpha\phi_n\\ \beta \phi_{n+1}
\end{array}\right) \,,\qquad n\geq 0\,, \label{psiLL}
\end{eqnarray}
with energy
\begin{eqnarray}
E_n=-\frac{\hbar\omega_c}{2} \pm \sqrt{
\left[\left(n+\frac 1 2 \right)\hbar\omega_c - \frac{(\hbar k_0)^2}{2m}
\right]^2 + 2 (n+1) \hbar\omega_cmv^2
}\,.\nonumber\\ \label{sqrtn}
\end{eqnarray}
Additionally, there is also the  ``$0$-LL'' state
\begin{eqnarray}
\psi_0 =e^{i\kappa y}\left( \begin{array}{c}
0 \\ \phi_{0}
\end{array}\right)\,,\\
\bar E_0=\frac{(\hbar k_0)^2}{2m} - \frac{\hbar\omega_c}{2}\,.\label{e0}
\end{eqnarray}
This eigenstate lies high above the Fermi level and is, therefore, empty.

We thus find that the spectrum contains an odd number of LLs: the Fermi level for
the half filled system
is a half-filled LL with high index $n=N$ that minimizes the square root in equation
(\ref{sqrtn}) and with energy above the loop gap
\begin{eqnarray}
E_N\approx -\frac{\hbar\omega_c}{2} + \Delta\,,
\end{eqnarray}
and $\alpha \approx \beta$ in the expression (\ref{psiLL}).

Had we chosen a Hamiltonian with opposite chirality, hence C=1, the $0$-LL state
would live on the other sublattice and have energy
symmetric to that in equation (\ref{e0}). So, it would be occupied.
The Fermi level would sit {\it below} the loop gap,
$E_N\approx \frac{\hbar\omega_c}{2} - \Delta$ which would be half-filled.
Therefore, the position of the Fermi LL with respect to the gap
 is the same as in the single Dirac cone problem\cite{eu}
\begin{eqnarray}
E_N\approx \left(\frac{\hbar\omega_c}{2} - \Delta\right) {\rm sgn}(C\cdot B)\,,
\end{eqnarray}
However, unlike the Dirac cone problem, where one has to consider at least two cones
because of the fermion doubling theorem\cite{NN}, we here may have only
one nodal loop in the BZ and get an odd number of LLs.

As before, each filled LL contributes $e^2/h$ to the charge Hall conductance, $\sigma_{yx}$.
The existence of the 0-LL either above or below the loop gap,
depending on the nodal loop's chirality,  determines the
Hall conductance in the gap. If $C=\pm1$, the 0-LL lies below/above and
$\sigma_{yx}=\pm e^2/h$ in the gap.

Endowing the electrons with spin, the simplest topological insulator\cite{Z2Kane}
with a Dirac loop gap and conserved $s_z$
could have a $C=1$  Hamiltonian for up spin electron and $C=-1$ for down spin.
The Chern matrix\cite{sc} would be diagonal with $C_\up=-C_\dn=1$.
The spin Chern number\cite{sc},
$C_{sc}=C_\uparrow -C_\downarrow = 2C$, with
$\nu=(C_{sc}\mathtt{mod}4)/2$.\cite{QWZtheorem,nagaosa07}
 The magnetic field breaks TRS, restoring the  $\mathbb{Z}$ index,  $C$, which counts
the number of edge states for each spin projection running in a given edge.
In thermal equilibrium the electrons migrate to the $E_N$ level sitting below the gap,
Eq.(\ref{e0}), so the system becomes spin polarized with spin density $|eB|/h$.
This is because the spin $\up$ electrons fill up their $0$-LL while the
spin $\dn$ electrons have it empty.
This spin density is half of that for two Dirac points, discussed in Ref\cite{eu}.
Note that the spin polarization is achieved without considering the Zeeman coupling to spin.
The  charge Hall conductance, $\sigma_{yx}=0$ because the two subsystems's
contributions  cancel. The spin Hall conductance, $\sigma_{yx}^z=e/(2\pi)$, however.
Such a  state is a spin Hall insulator with magnetization and it
is stable against potential disorder,
but unstable against  spin-flip perturbations, in which case it
would  become a trivial insulator.
\begin{figure}
\includegraphics[width=1\linewidth]{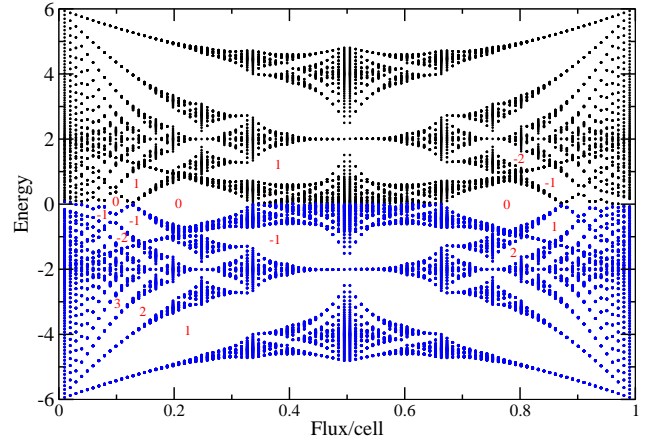}
\caption{(color online). Weyl loop Hofstadter spectrum of model  (\ref{model1}), which has
$C=-1$ at zero field. The flux per lattice cell is expressed in units of
the flux quantum,  $\phi_0=h/|e|$.
The occupied states for half filling are colored in blue. The Hall conductance
values
(in units of $e^2/h$) in some of the gaps is shown in red.}
\label{hofstadter1}
\end{figure}

\begin{figure}
\includegraphics[width=1\linewidth]{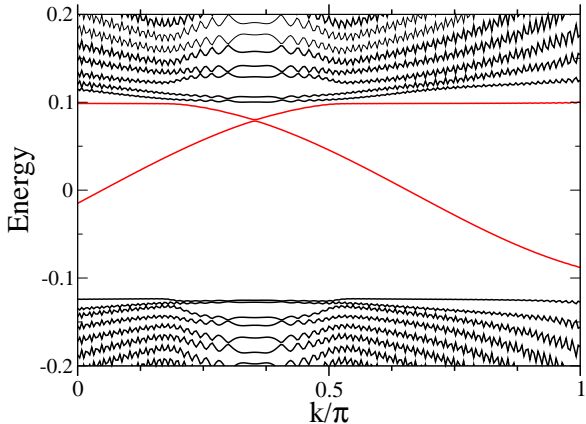}
\caption{(color online). Spectrum of model (\ref{model1}) for
a ribbon geometry under a weak magnetic field flux $\phi/\phi_0=1/400$,
vs. longitudinal momentum.
The edge states are highlighted in red. }\label{ribbon1}
\end{figure}

At stronger  magnetic field, the magnetic length
becomes comparable to the lattice spacing and the energy spectrum exhibits
the fractal structure known as the Hofstadter butterfly\cite{hofstadter}.
A calculation of Hofstadter butterfly spectra for Weyl nodes
in 3D systems has recently been done\cite{Roy}.
Fig. \ref{hofstadter1}
shows the Hofstadter spectrum for the model (\ref{model1})
with $\mu=2$, $\lambda=-0.1$ and $M=0$, at half-filling.
The Fermi energy lies above the gap, for small flux,
as the model's Chern number $C=-1$, in agreement with the above discussion.
Quantized Hall conductances (in units of $e^2/h$)
 are also shown in some of the Hofstadter gaps.
Although the half-filled system is metallic for small field,
it may become an insulator with zero Hall conductance, at higher flux values.

The spectrum for a ribbon geometry is shown in Fig. \ref{ribbon1}
for the same model, for a small magnetic flux.
Fig. \ref{ribbon1} confirms the  presence of edge states
crossing the gap with the predicted chirality.
An interesting difference between such a ribbon spectrum for a loop
and that of a Hamiltonian with two Dirac points\cite{eu} is readily
apparent. In the latter, the LL's dispersion with longitudinal momentum are
easily recognizable as plateaus, while in Fig. \ref{ribbon1} such LL plateaus are not seen.
The reason for this difference lies in the fact that LLs near the loop gap have high order,
so that the wave functions $\phi_n$ in
Eq.(\ref{LLsexpress}) contain high order Hermite polynomials.
The edge states therefore decay fairly slowly into
the bulk and finite size effects (the ribbon's width) are relatively strong.

\section{Conclusion}\label{conclsec}

We studied a minimal  Dirac ring Hamiltonian with mass gap terms,
for two-dimensional fermions,
as a model for  a Dirac loop anomalous Hall insulator.
We derived and expression for the Chern number which assigns a chirality
to the gapped loop through the phase winding of the mass gap terms.
The change in the Chern number at a topological transition can also be
calculated from a previously introduced index  that we generalized to the
Dirac loop case.

The Landau level spectrum in a weak magnetic field was shown to depend on the
loop's chirality.
The Fermi level has a high LL index.
In the  case of an
anomalous Hall insulator, a weak  magnetic field turns the system into a metal,
although it may become a trivial insulator at higher fields.
In the spinfull case of a topological insulator where spin $s_z$ is conserved,
the weak magnetic field's gauge field
turns the system into a spin Hall insulator with finite magnetization.

We also studied the  Hofstadter butterfly spectrum for arbitrary field, as well
as the edge states in a ribbon geometry.
The latter decay slower into the bulk and are therefore
more sensitive to finite size effects in the Dirac loop case,  when compared
to the case of Dirac nodes.

A recent proposal for the realization of a Dirac loop semimetal
in two dimensions has appeared recently\cite{Lu}.
We also expect that the models introduced above
are suitable for realization in optical lattices.

\end{document}